\documentclass{SSBAtran}

\usepackage[utf8]{inputenc}

\usepackage{graphicx}

\begin{document}
%
\title{Deep regression for uncertainty-aware and interpretable analysis of large-scale body MRI} 

%
%
%
%
\author{\IEEEauthorblockN{%
Taro Langner\IEEEauthorrefmark{1},
Robin Strand\IEEEauthorrefmark{1}\IEEEauthorrefmark{2},
H\r{a}kan Ahlstr\"{o}m\IEEEauthorrefmark{1}\IEEEauthorrefmark{3} and
Joel Kullberg\IEEEauthorrefmark{1}\IEEEauthorrefmark{3}}
\IEEEauthorblockA{\IEEEauthorrefmark{1}Department of Surgical Sciences, Uppsala University, 751 85 Uppsala, Sweden\\
Email: taro.langner@surgsci.uu.se}
\IEEEauthorblockA{\IEEEauthorrefmark{2}Department of Information Technology, Uppsala University, 751 05 Uppsala, Sweden}
\IEEEauthorblockA{\IEEEauthorrefmark{3}Antaros Medical AB, BioVenture Hub, 431 53 M\"{o}lndal, Sweden}
}

\maketitle

\begin{abstract}
Large-scale medical studies such as the UK Biobank examine thousands of volunteer participants with medical imaging techniques. Combined with the vast amount of collected metadata, anatomical information from these images has the potential for medical analyses at unprecedented scale. However, their evaluation often requires manual input and long processing times, limiting the amount of reference values for biomarkers and other measurements available for research.
Recent approaches with convolutional neural networks for regression can perform these evaluations automatically. On magnetic resonance imaging (MRI) data of more than 40,000 UK Biobank subjects, these systems can estimate human age, body composition and more. This style of analysis is almost entirely data-driven and no manual intervention or guidance with manually segmented ground truth images is required. 
The networks often closely emulate the reference method that provided their training data and can reach levels of agreement comparable to the expected variability between established medical gold standard techniques. The risk of silent failure can be individually quantified by predictive uncertainty obtained from a mean-variance criterion and ensembling. Saliency analysis furthermore enables an interpretation of the underlying relevant image features and showed that the networks learned to correctly target specific organs, limbs, and regions of interest.

\end{abstract}

\section{Introduction}

With 100,000 volunteer participants, the ongoing UK Biobank Imaging Study acquires vast volumes of medical imaging data \cite{sudlow_uk_2015, littlejohns2020uk}. Additionally, a wide range of metadata from other sources is collected, such as anthropometric measurements, biochemical assays, genetic information, and health outcomes. Magnetic resonance imaging (MRI) furthermore enables quantifications relating to body composition \cite{west_feasibility_2016} and liver fat \cite{wilman2017characterisation}. While the link between these properties and the image data is of great interest to medical research, their measurement often relies on manual input by trained image analysts. With quality control, placement of regions of interest and correction of semi-automated segmentation results, the evaluation of the image data poses a considerable challenge in time and labor. At the time of writing, reference values are therefore only available for a fraction of imaged subjects.
The large amount of image data allows for the training of convolutional neural networks for image-based regression. Similar approaches have been previously applied to dedicated MRI to estimate human age \cite{cole_brain_2018} and various properties of the heart \cite{xue2017direct}.

In this work, regression with convolutional neural networks on UK Biobank neck-to-knee body MRI is examined, with a closer look at recently presented approaches for estimates of age \cite{Langner2019}, general biometry \cite{Langner2020}, liver fat \cite{langner2020large}, and body composition \cite{langner2021uncertainty}. Whereas these publications provide extensive medical context, this submission aims to provide an overview and motivate the underlying design decisions with ablation experiments and further discussion of a more technical nature.

\section{Methodology}

\subsection{Segmentation and regression}

In biomedical image analysis, convolutional neural networks are commonly applied to perform measurements based on image segmentation. Architectures such as the U-Net \cite{ronneberger2015u} can be trained on carefully prepared ground truth segmentation images of dozens or hundreds of subjects to automatically label structures and tissues in two-dimensional representations of medical imaging data. In UK Biobank body MRI, related techniques have been proposed for segmentation of various organs \cite{bai2018automated, langner2020kidney, basty2020automated, bagur2020pancreas, irving2017deep}, muscles \cite{fitzpatrick2020large}, and other tissues \cite{liu2020systematic}.

The focus of this work is on regression with convolutional neural networks, or deep regression, which is methodologically distinct from segmentation techniques. Instead of training on pairs of images and ground truth segmentation masks, training samples for deep regression as discussed here consist of an input image and one or more numerical target values. The network output is consequently not a segmented image, but one or more continuous variables. 

Several desirable properties arise from this approach. Without requiring model-based assumptions, handcrafted features, or even direct access ground truth segmentation images for training, deep regression eliminates almost any need for manual intervention and guidance. Furthermore, abstract numerical properties can be inferred that could not otherwise be posed as a segmentation or classification task. Although, in general, more training samples are required, the implementations discussed here combine these advantages with high speed and accuracy and provide uncertainty estimates and saliency maps for quality control and interpretation.

\subsection{Deep regression on body MRI}

This work is focused on recent developments in image-based deep regression on UK Biobank neck-to-knee body MRI. On two-dimensional representations of the volumetric MRI data, variations of this technique can estimate human age \cite{Langner2019} and infer a wide range of medically relevant properties  ranging from features such as sex, height and weight to circumferences of waist and hip, volumes of muscle and adipose tissue depots as well as fat infiltrations \cite{Langner2020, langner2020large}. The underlying image features have been visualized, aggregated, and interpreted on a cohort level by saliency analysis with guided gradient-weighted class activation maps \cite{selvaraju_grad-cam:_2017}. More recently, a further extension for automated quality control has been proposed that uses a mean-variance loss function \cite{nix1994estimating} and ensembling \cite{lakshminarayanan2017simple} to provide predictive uncertainty as an estimate of individual prediction errors \cite{langner2021uncertainty}.
Although the underlying data can only be shared via UK Biobank access, code samples, documentation, and resulting saliency maps are publicly available.\footnote{https://github.com/tarolangner/mri-biometry}

\begin{figure}[t]
	\centering
	\includegraphics[width=0.95\columnwidth]{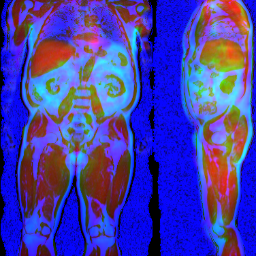}
	\caption{Two-dimensional representation of UK Biobank neck-to-knee body MRI for one subject, showing the water (red) and fat (green) signals as projections, together with noisy fat fraction values (blue) from two angles with \mbox{$256 \times 256 \times 3$} pixels.}
	\label{fig_input}
\end{figure}

\subsubsection{UK Biobank image data}

MRI can acquire volumetric representations of human anatomy without any known harmful side effects. Instead of ionizing radiation, magnetic fields are used to excite atomic nuclei in the imaged tissue, inducing electric currents in receiver coils that can be encoded as voxel-wise signal values. The UK Biobank neck-to-knee body MRI \cite{west_feasibility_2016, littlejohns2020uk} can be performed within about 6 minutes and is able to distinguish between signal obtained from water and fat molecules. As a result, volumetric water and fat signal images for more than 40,000 subjects have so far been released, with several overlapping stations that cover the body roughly from neck to knee, typically excluding the arms.

\subsubsection{Image formatting}

For the deep regression experiments, the separate stations of UK Biobank neck-to-knee MRI of each subject were first fused into a volume of \mbox{$370 \times 224 \times 174$} voxels, separately for both the water and fat signal. These volumes were then heavily compressed by mean intensity projection, forming the normalized sum of all coronal and all sagittal slices. By downsampling and combining this representation of both the water and the fat signal as color channels, a single, two-dimensional 8bit image of \mbox{$256 \times 256 \times 2$} pixels is obtained. A third channel can be added by extracting slices with voxel-wise fat fraction values, as shown in Fig. \ref{fig_input}.

\subsubsection{Neural network configuration}

Each subject forms one training sample, with the two-dimensional image format of Fig. \ref{fig_input} as input and one or more numerical values as regression targets. For training, the ground truth for these values was obtained from the UK Biobank metadata, based on reference measurements by techniques such as atlas-based segmentation \cite{west_feasibility_2016} or manual analyses \cite{wilman2017characterisation} with often incomplete coverage of the cohort.

As convolutional neural network architectures, both the VGG16 \cite{simonyan_very_2014} and the ResNet50 \cite{he_deep_2016} have been applied for this task. The target values were standardized and the network weights adapted from a model pretrained for classification on ImageNet. With a batch size of 32, online augmentation by random translations of up to 16 pixels, and the Adam optimizer, the network was trained in PyTorch with a base learning rate of 0.0001. After 5,000 iterations, the learning rate was reduced by factor ten, and training continued for another 1,000 iterations. 


\subsubsection{Uncertainty prediction}

The network can be configured to predict a single value for each subject and target when a mean squared error loss criterion is used. Unlike an output segmentation mask, however, the numerical output value alone provides little evidence for how plausible the prediction is, raising the risk of silent failure.
As an alternative approach, the loss function can be modified to accept two values. By modeling the mean and variance of a Gaussian probability distribution over each individual measurement, the optimization can be based on a maximum likelihood criterion \cite{nix1994estimating}. This approach poses little overhead and provides the output variance as estimate of uncertainty, with large values implying a potentially inaccurate prediction. When several networks of this type are combined into ensembles \cite{lakshminarayanan2017simple}, predictive uncertainty is obtained, which can identify some of the highest prediction errors automatically \cite{langner2021uncertainty}.

\subsubsection{Saliency analysis}

Guided gradient-weighted class activation maps can highlight those image areas of a specific input image that had a high impact on the network prediction \cite{selvaraju_grad-cam:_2017}. By co-aligning the anatomy of several subjects with image registration \cite{ekstrom_fast_2018}, these saliency maps can be aggregated for hundreds or thousands of subjects, yielding a cohort-wide visualization \cite{Langner2019}. These aggregated saliency maps enable an interpretation of the underlying decision criteria employed by the network for inference. 

\section{Results and discussion}

This section summarizes the results achieved by the regression networks, with focus on the generalized configuration \cite{Langner2020} and the more recent uncertainty-aware extension \cite{langner2021uncertainty}, and presents new insights and observations from ablation experiments.

\subsubsection{Predictive performance}

\begin{figure}[t]
	\includegraphics[width=\columnwidth]{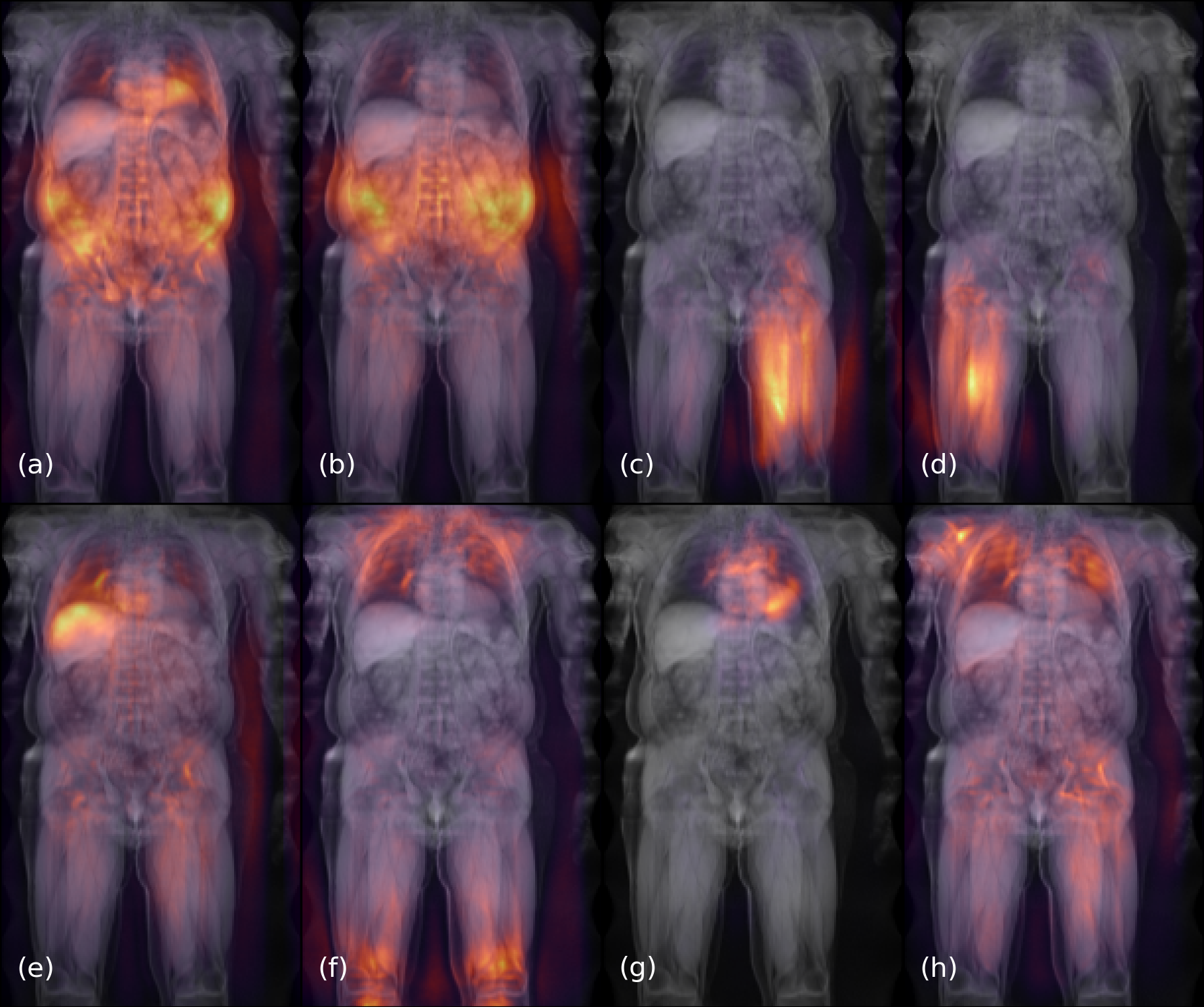}
	\caption{Aggregated ResNet50 saliency of about 3,000 subjects for: visceral adipose tissue as derived from atlas-based MRI segmentations (a)	or DXA \cite{littlejohns2020uk} (b), muscle volumes of the anterior left (c) and right thigh (d), liver fat (e), BMI (f), pulse rate (g) and grip strength (h). The network appears to emulate regions of interest used by different modalities and correctly targets specific limbs and organs \cite{Langner2020}.}
	\label{fig_sal}
\end{figure}

The deep regression techniques reach an accurate fit for several targets, reaching relative errors below 5\% or even 2\% on multiple measurements of body composition \cite{Langner2020, langner2021uncertainty}. Some measurements, such as volumes of adipose tissue depots in the abdomen, are available from two or more alternative reference techniques. Due to including varying regions of interest, these reference methods do not produce perfectly equal measurements, so that their agreement can be examined as a baseline. Based on the vast image data and numerical target values alone, the network learned to emulate these methods and inferred values that more faithfully reproduced their results than could be obtained by matching the reference methods to each other with linear transformations. The saliency analysis indicates that the network also learned to mimic the regions of interest and correctly targets individual organs and limbs, as seen in Fig. \ref{fig_sal} \cite{Langner2020}. Not only did the quantification of thigh muscle lead to salient features being located in the correct leg, but the prediction also reached an accuracy that exceeded the natural similarity of both legs to each other \cite{Langner2020}. All of these results were achieved with a standardized training policy \cite{Langner2020}.

With a Nvidia RTX 2080 Ti 11GB on a desktop computer with 32GB RAM, network training completed a single split with 6,000 iterations in just about 15 minutes. Once all input images are cached, predictions for 30,000 subjects required less than 10 minutes even when ensembling was used.


\subsubsection{Quantity of training data}

\begin{figure}[t]
	\includegraphics[width=\columnwidth]{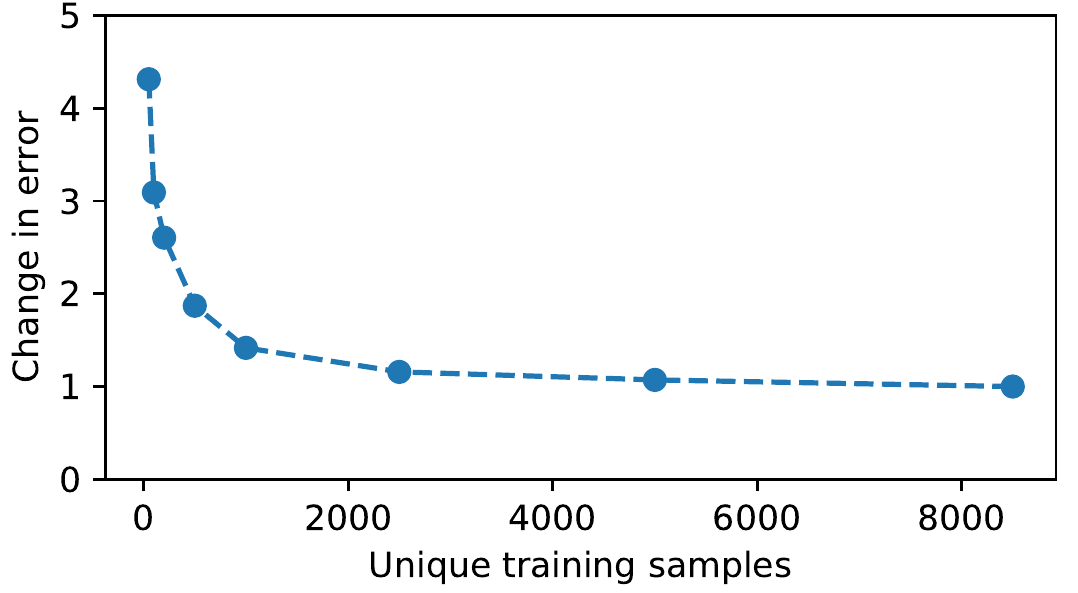}
	\caption{Effects of limited training data. A mean-variance regression network was trained repeatedly with varying amounts of unique training samples to predict six body composition targets on one split of a previous cross-validation experiment \cite{langner2021uncertainty}. The y-axis represents the average change in the mean absolute percentage error (MAPE). The baseline error, as achieved with 8,500 training samples, is accordingly doubled when less than 450 samples are available, tripled with less than 120 samples, and quadrupled with less than 60 samples. The exact numbers and performance vary depending on the target (see also \cite{Langner2019}).
  }
	\label{fig_training_count}
\end{figure}

For segmentation with neural networks, successful training on UK Biobank neck-to-knee body MRI has been reported with annotated data of between 90 and 220 subjects \cite{fitzpatrick2020large, bagur2020pancreas}. Each subject can effectively supply multiple training sample in the form of two- or three-dimensional patches. The image-based regression techniques discussed here only obtain one unique training sample for each subject, and therefore require a larger number of subjects for training. As seen in Fig. \ref{fig_training_count}, the error generally decreases in exponential decay along with a rising number of unique training samples.

\subsubsection{Image compression}

Aggressive compression to a two-dimensional 8bit input format was required for viable speed and memory usage. The mean intensity projections reduce the number of values obtained from MRI to just about $0.5\%$. The networks were nonetheless able to accurately infer circumferences, volumes and even abstract properties outside of the field of view from these images, such as thigh muscle volume and grip strength. However, including fat fraction slices in the input reduced the absolute prediction error for liver fat roughly by half \cite{Langner2020, langner2020large}. Despite the empirical success on many targets, the image compression accordingly causes a loss of relevant information. On a conceptual level, the choice of formatting resembles a selection of handcrafted features, and it would be desirable to include the compression in an end-to-end learning procedure for joint optimization. It is likely that this would require substantially more powerful hardware and specialized volumetric network architectures.




\subsubsection{Standardization of target labels}
Standardization for removal of scaling and offsets in the target label greatly increased training stability and convergence. While the VGG16 architecture \cite{simonyan_very_2014} was able to directly learn the prediction of age in days \cite{Langner2019}, architectures with skip connections only became stable when the target value was standardized by subtracting its mean and dividing by its standard deviation as a pre-processing step, and reversing these steps on the network output as post-processing. 


\subsubsection{Architecture selection}

\begin{figure}[t]
	\includegraphics[width=\columnwidth]{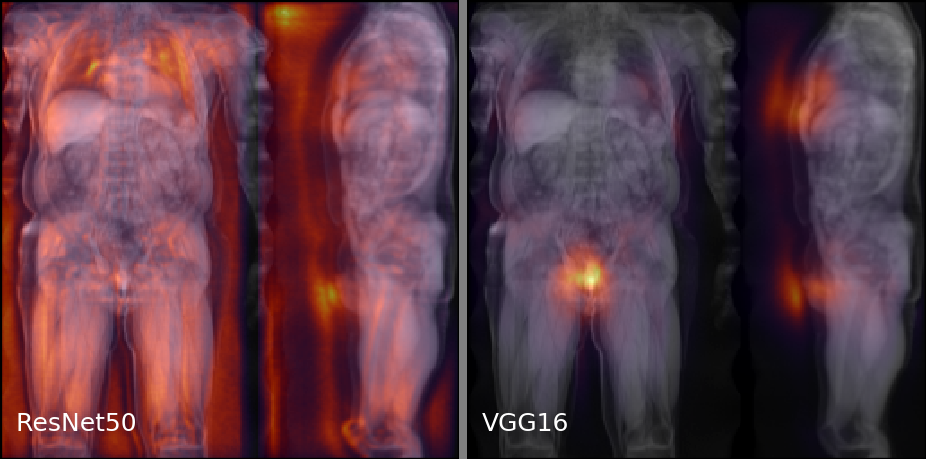}
	\caption{Aggregated saliency of about 3,000 subjects in prediction of sex. Both networks identify women or men with an accuracy $\ge99,97\%$, but whereas the ResNet50 is diffuse, the VGG16 saliency is strikingly plausible and easily supported by anecdotal evidence.}
	\label{fig_sal_compare}
\end{figure}

Target value standardization and initialization with ImageNet-pretrained weights appears to enable the VGG16 with batch normalization \cite{simonyan_very_2014}, ResNet50 \cite{he_deep_2016}, DenseNet161 \cite{huang_densely_2017}, and InceptionV3 \cite{szegedy2016rethinking} to perform roughly on par. Variations of the EfficientNet \cite{tan2019efficientnet}, InceptionV4 and Inception-ResNet \cite{szegedy2017inception} showed slightly worse performance in this configuration. The ResNet50 runs about $17\%$ faster than the VGG16, uses less GPU memory, and stores snapshots only $14\%$ of its size. However, it was found to produce more diffuse saliency maps, as seen in Fig. \ref{fig_sal_compare}, possibly due to its residual blocks and global average pooling layer.

\subsubsection{Robustness of the configuration}
\begin{figure}[t]
	\includegraphics[width=\columnwidth]{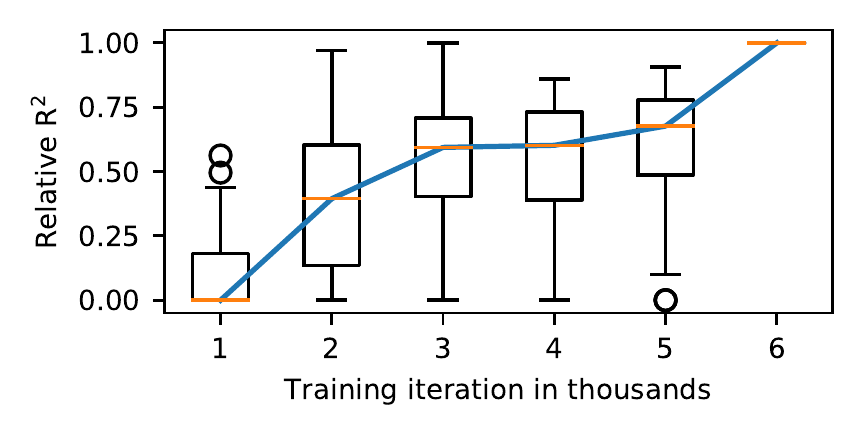}
	\caption{Training stability of the biometry configuration on 64 targets \cite{Langner2020}. Each box represents the spread of cross-validation scores, aggregated for all targets, by one of the six snapshots saved after 1,000 training iterations each. Box extents cover the lower to upper quartile of scores, with an orange line at the median, whiskers at 1.5 times the interquartile or actual range, and outliers beyond as circles. All networks reached their highest R$^2$ at the final snapshot.}
	\label{fig_train_curves}
\end{figure}

The network configuration for biometry consistently reduced the validation error in training and avoided overfitting for all 64 fields \cite{Langner2020}. The same training policy was used for the more recent, uncertainty-aware approach \cite{langner2021uncertainty}, with similar results. On large sample sizes of the UK Biobank, repeated validation yields virtually the same average performance, sometimes with superior results on test data after learning from all available training samples. The final reduction of the learning rate helps to stabilize training, as visualized in Fig. \ref{fig_train_curves} with aggregate learning curves.

\subsubsection{Uncertainty prediction}

When mean-variance loss and ensembling are combined, estimates of predictive uncertainty for each individual prediction can be provided. High uncertainty can indicate potential failure cases and has been observed in anomalous cases, such as an inaccurate prediction of thigh muscle in a subject with an atrophied right leg or image artifacts \cite{langner2021uncertainty}. 

Apart from requiring a second output value, the mean-variance regression imposes no overhead and did not negatively affect runtime, memory requirements, or convergence. In return, it achieved slightly superior accuracy, likely as a result of \textit{loss attenuation}, as outliers in the ground truth can be accounted for by high variance alone \cite{kendall2017uncertainties}.
Ensembling requires several network instances to be trained, with each directly contributing to runtime requirements. However, there are diminishing returns to increasing the ensemble size, and smaller ensembles with just five instead of ten network instances may be sufficient. Regardless of the chosen loss function, ensembling also benefits overall prediction accuracy.

High uncertainty can identify some of the worst individual prediction errors, but was also noted to correlate with high-valued predictions. Accordingly, heavyweight subjects may be mistakenly excluded due to inherently higher uncertainty and future work may be required to more clearly separate these two effects.

\subsubsection{General limitations}

The trained networks can only be expected to generalize to future UK Biobank neck-to-knee body MRI with the same imaging protocol, demographics (males and females aged 44-82 years, 95\% self-reported white British ethnicity) and quality-controlled images. Deviation from these constraints may require retraining on new data. It is also worth noting that not all properties can be inferred in the presented way, and metadata related to smoking, liver iron content and inflammation, blood pressure, and blood cholesterol could not be inferred with satisfactory accuracy as of yet. It is possible that 
information relating to these properties is contained in the image data and could be leveraged by more suitable image formats to be examined in future work.

\section{Conclusion}

Deep regression with convolutional neural networks has the potential for fast and accurate image-based inference of biological measurements, with minimal need for human intervention or guidance. Predictive uncertainty can highlight potential failure cases and aggregated saliency analysis can enable an interpretation of the underlying, relevant image features. By deploying these frameworks, missing UK Biobank metadata can be conveniently inferred on a large scale for further medical studies, quality control, and genetic research.



\section*{Acknowledgment}
This research was supported by a grant from the Swedish Heart-Lung Foundation and the Swedish Research Council (2016-01040, 2019-04756, 2021-70492) and used the UK Biobank Resource under application no. 14237. The authors would like to thank Fredrik K. Gustafsson and Benny Avelin for their advice on ensembling, probabilistic modeling, and uncertainty estimation.




\bibliographystyle{SSBAtrans}
\bibliography{sample}

\end{document}